\documentclass[twocolumn,superscriptaddress,prl,10pt]{revtex4-2}

\usepackage{amsmath,amssymb,mathtools}
\usepackage{xcolor}
\usepackage{graphicx}
\usepackage{bm}

\newcommand{\bE}{\mathbb{E}}
\newcommand{\UHP}{\text{UHP}}
\newcommand{\bR}{\mathbb{R}}
\newcommand{\bC}{\mathbb{C}}

\newcommand{\cN}{\mathcal{N}}

\usepackage{dsfont}

\begin{document}

\title{String Theory from Infinite Width Neural Networks}

\author{Samuel Frank}
\affiliation{Department of Physics, Northeastern University, Boston, MA 02115 USA}

\author{James Halverson}
\affiliation{Department of Physics, Northeastern University, Boston, MA 02115 USA}
\affiliation{The NSF AI Institute for Artificial Intelligence and Fundamental Interactions}

\begin{abstract}
We realize bosonic string theory with ensembles of infinite width neural networks. The string tension is tuned by the variance of the output weights.  The construction provides a new computation of the foundational Virasoro-Shapiro and Veneziano amplitudes as neural network  correlators. 
\end{abstract}

\maketitle

\section{Introduction}

Recent advances in machine learning (ML) theory have motivated an approach to field theory that relies crucially on the defining data of neural networks. 

In this Letter we extend the correspondence to quantum gravity by realizing the bosonic string with ensembles of neural networks. We do so by specifying network architectures and probability densities on their parameters, chosen to reproduce the statistics of the free boson and $bc$-ghost system. This description enables a new computation of the Veneziano and Virasoro-Shapiro amplitudes in network parameter space, recovering essential results from string theory that exhibit crossing symmetry, Regge behavior, and softness in the ultraviolet.

The neural network field theory (NN-FT) approach to field theory \cite{halverson2021building} builds on
an essential ML result \cite{neal} that many deep neural networks (NN) admit large-$N$ limits in which they are draws from Gaussian processes, enabling a theoretical treatment in terms of free field theory. This is the neural network Gaussian process (NNGP) correspondence. The functional form of the network is known as the architecture, and $N$ is a measure of its size. For instance, $N$ is the width of a feedforward network, the number of channels in a convolutional neural network \cite{Novak2018BayesianCN, GarrigaAlonso2019DeepCN}, or the number of attention heads \cite{hron2020infinite} in the attention mechanism  \cite{vaswani2017attention} that has led to breakthroughs in language models. Many other architectures have been studied \cite{williams, Matthews2018GaussianPB, schoenholz2017correspondence, Jacot2018NeuralTK}; see \cite{yangTP1, yangTP2} for a unified perspective.  Nevertheless, networks encountered in the wild have finite-$N$, and accordingly have interactions captured by non-Gaussianities \cite{Yaida2019NonGaussianPA, PhysRevE.104.064301, Halverson_2021, roberts2022principles}.

In \cite{halverson2021building} it was proposed to use these techniques for field theory, rather than for ML.  A number of results are known in this approach, including the origin of symmetries \cite{maiti2021symmetryviaduality} and interactions \cite{ Halverson_2021,demirtas2023neural}; the relation to Euclidean \emph{quantum} field theories  and Euclidean-invariant architectures \cite{halverson2021building}; a realization of the free scalar \cite{halverson2021building}, free Dirac spinor \cite{Frank:2025zuk}, $\phi^4$ theory \cite{demirtas2023neural}, and Yukawa interactions \cite{Frank:2025zuk}; and  new interacting supersymmetric theories \cite{Frank:2025zuk}. Neural network conformal field theories were introduced in \cite{Halverson:2024axc} and conformal defects in \cite{Capuozzo:2025ozt}, relying on the embedding formalism and not restricting the dimension.

Very recently, Robinson \cite{Robinson:2025ybg} introduced NN-FTs that realize the full Virasoro symmetry of 2d CFT in the infinite width limits of certain neural networks. Other classic results include NN-FT realizations of the 2d free boson, the free fermion, and an $\cN=(1,1)$ multiplet.
This work and \cite{Robinson:2025ybg} are largely complementary, but do have some overlapping results. We thank the author for sharing his manuscript while ours was being finalized.

We will develop NN-FT descriptions of the 2d free boson CFT and the $bc$-ghost system. Together, these determine the gauge-fixed Polyakov action of bosonic string theory. We then utilize the new description to compute well-known string amplitudes. 

\section{The Bosonic String as a NN-FT}

We realize the bosonic string as a NN-FT by defining architectures and parameter densities that yield the free boson and $bc$-ghost system in the infinite width limit.

\medskip
\noindent \textbf{Free Boson}.
 We choose an architecture to represent the closed bosonic string embedding $X^\mu(z,\bar z)$ in $D$-dimensional Euclidean spacetime. The Euclidean worldsheet coordinates are $(x_1,x_2) \in \bR^2$, but we work in complex coordinates with $z=x_1+ix_2$ and $\bar{z}=x_1-ix_2$. The architecture is 
\begin{equation}
    X^\mu(z)=\frac{C}{\sqrt{N}}\sum_{i=1}^N \frac{a_i^\mu}{|W_i|}\cos\Big(\frac{1}{2}(W_i z+\bar{W}_i\bar{z})+c_i\Big),\label{eq:X^mu_NNFT}
\end{equation}
where $\mu=1,\ldots,D$ labels spacetime directions and the factor $C=\sqrt{\left(\alpha' (\Lambda^2-\epsilon^2)\right)/\sigma_a^2}$ normalizes the two-point function. 
The parameters are distributed i.i.d. as 
\begin{align}
    a_i^\mu &\sim \mathcal{N}(0, \sigma_a^2) \notag\\
    c_i &\sim \text{Uniform}[-\pi,\pi] \notag\\
    W_i &\sim \text{Uniform}(B^2_\Lambda\backslash B^2_\epsilon), \quad W_i \in \mathbb{C}\label{eq:param_dist_X}
\end{align}
where $B^2_\Lambda\backslash B^2_\epsilon$ denotes the 2d annulus in $\bC$ of inner radius $\epsilon$ and outer radius $\Lambda$. The hyperparameters $\epsilon$ and $\Lambda$ represent the IR and UV momentum cutoffs, respectively, while $\alpha'$ appears in the normalization to match the standard string theory convention. 

\emph{Correlators.}
We compute the two-point function $\langle X^\mu(z) X^\nu(w)\rangle$ by averaging over the parameter distributions. Taking expectations over $a$ and $c$ while using independence along the $i$ index yields
\begin{equation}
    \langle X^\mu(z) X^\nu(w)\rangle=\frac{\sigma_a^2 C^2}{2}\;\mathbb{E}_W\Big[\frac{\cos[(W\Delta z+\bar{W}\Delta\bar{z})/2]}{|W|^2}\Big] \delta^{\mu\nu},
\end{equation}
where $\Delta z:=z-w$ and $\Delta\bar{z}:=\bar{z}-\bar{w}$. Define $r:=|\Delta z|$. Then the expectation over $W$ becomes an integral involving the Bessel function $J_0$:
\begin{align}
\label{eqn:XX_pos}
    \langle X^\mu X^\nu\rangle(r)&=\alpha'\int_{\epsilon}^\Lambda d|W|\,\frac{J_0(r|W|)}{|W|} \delta^{\mu\nu}\notag\\
    &=\alpha'\delta^{\mu\nu}\Big[-\log r+\log\frac{2\Lambda}{\epsilon e^\gamma}+\dots\Big]\;,
\end{align}
where $\gamma$ is the Euler-Mascheroni constant. This recovers the standard bosonic string propagator in position space for $1/\Lambda\ll r \ll 1/\epsilon$.
In momentum space with Fourier conventions $F(p) = (2\pi)^2\int d^2x\, e^{-ip\cdot x} F(x)$:
\begin{equation}
    \langle X^\mu X^\nu\rangle(p)=\frac{2\pi\alpha'}{p^2} \delta^{\mu\nu} \bm{1}_{B^2_\Lambda\backslash B^2_\epsilon}\;,
\end{equation}
reproducing the characteristic $1/p^2$ scaling of the string propagator for $\epsilon < |p| < \Lambda$.
This correlator is the same for all $N$, but by the Central Limit Theorem (see, e.g., \cite{demirtas2023neural}) higher connected correlators vanish for $N\to\infty$, yielding the required free theory.

How should one understand the string length $\ell_s \propto \sqrt{\alpha'}$? The constant $C$ was carefully chosen with the form above to exactly realize the $\alpha'$ in \eqref{eqn:XX_pos}. Without this foresight, one might have instead chosen $C=1$, in which case matching the coefficient of the $\langle X X \rangle$ correlator to $\alpha'$ would yield
\begin{equation}
    \alpha' = \frac{\sigma_a^2}{\left(\Lambda^2 - \epsilon^2\right)}\;.
\end{equation}
The string length is determined in this case by the standard deviation $\sigma_a$ of the output weights and determines the tension as $T\sim 1/\ell_s^2$. The only (non-regulator) hyperparameter in the parameter densities, $\sigma_a$, determines the only parameter in the physical theory, $\alpha'$.

\emph{Symmetries and Zero Modes.} Following \cite{halverson2021building}, rotation and translation invariance on the worldsheet are manifest by the choice of distributions for $W_i$ and $c_i$. Since in string theory spacetime is the target space, symmetries in spacetime are associated with the outputs of the neural network. Spacetime rotational invariance is manifest due to the choice of rotationally invariant Gaussians for $a_i^\mu$. We include a zero mode as 
\begin{equation}
\tilde X^\mu(z) = X^\mu(z) + X_0^\mu\,, \qquad X_0^\mu \sim \cN(0,\sigma_0^2)\;,    
\end{equation}
which may be interpreted as adding a Gaussian bias to the architecture.
Under spacetime translations $\tilde X^\mu \mapsto \tilde X^\mu + C^\mu$ the translation may be absorbed into a parameter redefinition ${X_0^{\mu}}'  := X_0^\mu + C^\mu$. In the limit $\sigma_0\to \infty$ the density on $X_0^\mu$ is invariant under this transformation, which we will see is one way to obtain spacetime momentum conservation in string amplitudes. 

\medskip
\noindent \textbf{Ghosts}.
We choose the $bc$ ghost architectures to be
\begin{align}
    b(z,\bar{z})&=\frac{\Lambda}{\sigma}\frac{1}{\sqrt{2N}}\sum_{k=1}^N \frac{\beta_k}{\sqrt{|\omega_k|}}e^{i\big(\frac{1}{2}(\omega_k z + \bar{\omega}_i \bar{z})+c_k\big)}\notag\\
    c(z,\bar{z})&=\frac{\Lambda}{i\sigma}\frac{1}{\sqrt{2N}}\sum_{k=1}^N \frac{\chi_k}{\sqrt{|\omega_k|}}e^{-i\big(\frac{1}{2}(\omega_k z + \bar{\omega}_i \bar{z})+c_k-\phi_k\big)},\notag
\end{align}
following \cite{Frank:2025zuk} in defining Grassmann-valued neural networks.
The parameters are distributed i.i.d. as
\begin{align}
    \beta_k, \chi_k &\sim \text{Grassmann Gaussian}: \; P\propto\exp(\beta_k \chi_k/\sigma^2) \notag\\
    c_k &\sim \text{Uniform}[-\pi,\pi] \notag\\
    \omega_k &\sim \text{Uniform}(B^2_\Lambda), \quad \omega_k \in \mathbb{C}, \notag
\end{align}
 where $B^2_\Lambda$ denotes the 2d disk of radius $\Lambda$. The complex frequency $\omega_k=|\omega_k|e^{i\phi_k}$ is defined to have phase $\phi_k$.  The Grassmann weights satisfy $\mathbb{E}[\beta_k \chi_{l}]=\sigma^2\delta_{kl}$, $\mathbb{E}[\beta_k \beta_{l}]=0$, and $\mathbb{E}[\chi_k \chi_{l}]=0$, using Berezinian integration. 

Computing the two-point function by evaluating first the $\beta$, $\chi$, and $c$ expectations yields:
\begin{equation}
    \langle b(z) c(w)\rangle=\frac{\Lambda^2}{2i}\mathbb{E}_\omega\Big[\frac{e^{\frac{i}{2}(\omega\Delta z+\bar{\omega}\Delta \bar{z})}e^{i\phi}}{|\omega|}\Big]\;.
\end{equation}
Defining $\Delta z = re^{i\theta}$ allows us to write 
\begin{align}
    \langle b(z) c(w)\rangle&=e^{-i\theta}\int_0^\Lambda d|\omega|\,J_1(r|\omega|)
    = \frac{1-J_0(\Lambda r)}{re^{i\theta}}\notag\\[10 pt]
    &=\frac{1}{z-w}+\dots\;
\end{align}
with $\mathcal{O}((\Lambda r)^{-1/2})$ corrections vanishing as $\Lambda r\gg1$. Computing the Fourier Transform gives:
\begin{equation}
    \langle bc\rangle(p)=-\frac{2\pi i}{\bar{p}}\bm{1}_{B^2_\Lambda}\;,
\end{equation}
as desired for the $bc$-ghost system. The $\bar b \bar c$-system is realized by an independent copy $(\bar\beta_k,\bar\chi_k)$ of the Grassmann weights (uncorrelated with $(\beta_k,\chi_k)$), and the replacement $\phi_k\to-\phi_k$ in the $c$-architecture. This produces $\langle\bar b \bar c\rangle(\bar{z}, \bar{w})=1/(\bar{z}-\bar{w})+\dots$, with mixed barred-unbarred correlators vanishing.

\section{String Amplitudes}

Equipped with a new description of the bosonic string, we now utilize it to compute famous string amplitudes.

\medskip
\noindent \textbf{Virasoro-Shapiro Amplitude}.
The tree-level four-point scattering amplitude for closed string tachyons is:
\begin{equation}
    \mathcal{A}^{(4)}(p_1,\dots,p_4)=\frac{1}{g_s^2}\frac{1}{\text{Vol}(SL(2,\mathbb{C}))}\Big\langle \prod_{i=1}^4 V(p_i)\Big \rangle\;,\label{eq:VS_definition}
\end{equation}
where $g_s$ is a hyperparameter corresponding to the string coupling, and $V(p)=g_s\,\mathcal{Z}(p)\int d^2z\,e^{ip\cdot \tilde X(z)}$ is the vertex operator with $\mathcal{Z}(p)$ a renormalization factor whose importance will be seen in the following. Instead of evaluating the expectation with the gauge fixed Polyakov action, we instead utilize our NN description, integrating with respect to the measure \eqref{eqn:no_zero_mode_measure} together with the Gaussian measure of the zero mode; see Appendix for a full calculation. 

Evaluating the expectation of the zero mode $X_0$ and the output weights $a_i^\mu$, 
\begin{align}
    \mathcal{A}^{(4)}&(p_1,\dots,p_4)
    =\frac{g_s^2\,e^{-\sigma_0^2 (\sum_i p_i)^2/2}}{\text{Vol}(SL(2,\mathbb{C}))}\int \Big(\prod_{\alpha=1}^4 d^2z_\alpha\Big)\notag\\
    &\times\mathcal{Z}(p_1)\dots\mathcal{Z}(p_4)
    \exp\Big(-\frac{\alpha'}{2}(\Lambda^2-\epsilon^2)\,\mathbb{E}_{W,c}[Q]\Big)\;,
\end{align}
where we have used the large $N$ limit and used the i.i.d. property to remove the $i$ index. We have
\begin{align}
    Q &=\sum_{r,s=1}^4 (p_r\cdot p_s)f(z_r) f(z_s) \\[5 pt]
    f(z)& =  \frac{\cos\Big(\frac12 (W z + \bar W \bar z) + c\Big)}{|W|}\;.
\end{align}
Splitting the  sum and computing the $c$ expectation,
\begin{align}
    \mathbb{E}_{W,c}[Q]
    &=\mathbb{E}_W\Big[\frac{1}{2}\sum_{r\neq s}\,(p_r\cdot p_s)\,\frac{\cos\big(\frac12 (W \Delta z_{rs} + \bar W \Delta\bar z_{rs})\big)}{|W|^2}\notag\\\notag
    &\qquad\qquad\qquad+\frac{1}{2}\sum_r\,p_r^2\,\frac{1}{|W|^2}\Big]\;\\
    &=-\frac{1}{\Lambda^2-\epsilon^2}\sum_{r\neq s}\,(p_r\cdot p_s)\log|\Delta z_{rs}| + C(\Lambda,\epsilon,p)\;,
    \label{eq:expec_Q} 
\end{align}
where $\Delta z_{rs} = z_r-z_s$ and 
\begin{align}
    C(\Lambda,\epsilon,p)
    & = \frac{\Big(\sum_r p_r\Big)^2\,\log\Big(\frac{2\Lambda}{\epsilon e^\gamma}\Big) - \sum_r\,p_r^2\,\log\Big(\frac{2}{e^\gamma}\Big)}{\Lambda^2-\epsilon^2}
    \label{eq:cutoff_dep_const}\;.
\end{align}
We emphasize this splitting of $\bE_{W,c}[Q]$ into propagator pieces and cutoff-dependent pieces. The latter emerged from contractions of $X(z_r)$'s from a single vertex operator, which would disappear if we had utilized normally ordered vertex operators with $\mathcal{Z}(p)=1$. Instead, we have foregone normal ordering, choosing instead
\begin{equation}
    \mathcal{Z}(p)=\Big(\frac{e^\gamma}{2}\Big)^{\alpha'p^2/2} \left(\frac{\sigma^2_\text{eff}}{2\pi}\right)^{D/8}\;,
\end{equation}
where $\sigma_{\text{eff}}^2=\sigma_0^2+\alpha'\log(2\Lambda/\epsilon e^\gamma)$. This choice has two effects:
to cancel $p_r^2$-dependent terms in the limit, and to yield a  Gaussian normalization.
We have
\begin{align}
    \mathcal{A}^{(4)}&=\frac{g_s^2\left(\frac{\sigma^2_\text{eff}}{2\pi}\right)^{\frac{D}{2}}e^{-\frac{\sigma_{\text{eff}}^2(\sum_r p_r)^2}{2}}}{\text{Vol}(SL(2,\mathbb{C}))} \nonumber \\
    & \qquad \times \int\Big(\prod_{\alpha=1}^4 d^2z_\alpha\Big)\,\prod_{r<s}|z_r-z_s|^{\alpha'p_r\cdot p_s}\;,
\end{align}
and taking $\sigma_{\text{eff}}\to\infty$, we obtain 
\begin{align}
    \mathcal{A}^{(4)}&=\frac{g_s^2\,\delta^{D}(\sum_r p_r)}{\text{Vol}(SL(2,\mathbb{C}))}\int\Big(\prod_{\alpha=1}^4 d^2z_\alpha\Big)\,\prod_{r<s}|z_r-z_s|^{\alpha'p_r\cdot p_s}\;,
\end{align}
which is the Virasoro-Shapiro amplitude.  One may cast this in a form with $\Gamma$-functions by using conformal symmetry to fix $z_1=\infty$, $z_2=0$, $z_3=z$, $z_4=1$   and evaluating the remaining integral. Alternatively, one may compute the amplitude with $c$-ghost insertions and consideration of their zero modes on the sphere.

\medskip
\noindent \textbf{Veneziano Amplitude.}

\emph{Open String NN-FT.}
The Veneziano amplitude is the analog of the Virasoro-Shapiro amplitude for open strings. Unlike for closed strings, that were defined on $\mathbb{C}$, the worldsheet $\Sigma$ of open strings is chosen to be the upper half plane ($\UHP$) defined by $\Im z\geq0$. Since the $\UHP$ has a boundary $\partial\Sigma$ (the real axis), we must enforce a boundary condition on the open string NN-FT. We choose Neumann boundary conditions, 
\begin{equation}
    (\partial-\bar{\partial})X^\mu_N(z,\bar{z})\Big|_{z=\bar{z}}=0\;,\label{eq:neumann}
\end{equation}
expressed in complex coordinates, where $\partial:=\partial_z$, $\bar{\partial}:=\partial_{\bar{z}}$, and $X_N(z,\bar{z})$ is a modification of the closed string architecture \eqref{eq:X^mu_NNFT} that obeys these boundary conditions:
\begin{align}
    X^\mu_N(z,\bar{z})&=\frac{1}{\sqrt{2}}\big[X^\mu(z,\bar{z})+X^\mu(\bar{z}, z)\big]\notag\\
    &=\frac{C}{\sqrt{2N}}\sum_{i=1}^N\frac{a_i^\mu}{|W_i|}\Big[\cos\Big(\frac{1}{2}(W_i z+\bar{W}_i \bar{z})+c_i\Big)\notag\\
    &\qquad+\cos\Big(\frac{1}{2}(W_i \bar{z}+\bar{W}_i z)+c_i\Big)\Big]\;,
\end{align}
with $z\in\UHP$ and parameters are distributed as in \eqref{eq:param_dist_X}. 
Neumann boundary conditions follow from observing that $X^\mu_N(z,\bar{z})=X^\mu_N(\bar{z},z)$, so $\partial X^\mu_N(z,\bar{z})=\bar{\partial}X^\mu_N(\bar{z},z)$, satisfying the condition \eqref{eq:neumann} on $\partial\Sigma$.

 Let $G_{\mathbb{C}}(z,w)$ be the closed string two-point function \eqref{eqn:XX_pos}, then 
 $\langle X^\mu_N(z,\bar{z})\,X^\nu_N (w,\bar{w})\rangle = G_{\mathbb{C}}(z,w)+G_{\mathbb{C}}(z,\bar{w})$.
Restricting to the real axis, the open-string two-point function is 
\begin{align}
    \langle X^\mu_N(x)\,X_N^\nu(x')\rangle\Big|_{\partial\Sigma}&=2G_{\mathbb{C}}(x,x')\notag\\
    &=2\alpha'\delta^{\mu\nu}\Big[-\log r +\log\frac{2\Lambda}{\epsilon e^\gamma}+\dots\Big]\;,\label{eq:boundary_prop}
\end{align}
where
$x,x'\in\mathbb{R}$, and $r=|x-x'|$.

\emph{The Veneziano Amplitude.}
Now we turn to the calculation of the Veneziano amplitude, defined analogously to the Virasoro-Shapiro amplitude \eqref{eq:VS_definition}:
\begin{align}
    \mathcal{A}^{(4)}(p_1,\dots,p_4)=\frac{1}{g_s}\frac{1}{\text{Vol}(SL(2,\mathbb{R}))}\Big\langle\prod_{i=1}^4 V_o(p_i)\Big\rangle\;,
\end{align}
where $SL(2,\mathbb{R})$ is the subgroup of $SL(2,\mathbb{C})$ that preserves the boundary of the $\UHP$, and the vertex operator $V_o(p)=\mathcal{Z}_o(p)\sqrt{g_s}\int dx\,e^{ip\cdot \tilde{X}_N}$, with $\tilde{X}^\mu_N=X^\mu_N+X^\mu_0$ (the zero mode $X^\mu_0$ is the same as in the closed string case). The string embedding $\tilde X^\mu_N$ is restricted to the real axis.

The calculation of the Veneziano Amplitude proceeds similar to that of Virasoro-Shapiro, with the only substantive change coming from the doubled boundary two-point function \eqref{eq:boundary_prop} that causes several coefficients to be scaled by a factor of 2. The final result is
\begin{align}
    \mathcal{A}^{(4)}=&\frac{g_s\left(\frac{{\sigma'}^2_\text{eff}}{2\pi}\right)^{\frac{D}{2}}e^{-\frac{{\sigma'}_{\text{eff}}^2(\sum_r p_r)^2}{2}}}{\text{Vol}(SL(2,\mathbb{R}))}\notag\\
    &\qquad\times
    \int\Big(\prod_{\alpha=1}^4 dx_\alpha\Big)\,\prod_{r<s}|x_r-x_s|^{2\alpha'p_r\cdot p_s}\;,\label{eq:Venez_pre}
\end{align}
where ${\sigma'}_{\text{eff}}^2=\sigma_0^2+2\alpha'\log(2\Lambda/\epsilon e^\gamma)$, and the renormalization factors are
\begin{equation}
    \mathcal{Z}_o(p)=\Big(\frac{e^\gamma}{2}\Big)^{\alpha'p^2} \left(\frac{{\sigma'}^2_\text{eff}}{2\pi}\right)^{D/8}\;.
\end{equation}
In the $\sigma'_{\text{eff}}\to\infty$ limit, we recover
\begin{equation}
    \mathcal{A}^{(4)}=\frac{g_s\,\delta^{D}(\sum_r p_r)}{\text{Vol}(SL(2,\mathbb{R}))}\int\Big(\prod_{\alpha=1}^4 dx_\alpha\Big)\,\prod_{r<s}|x_r-x_s|^{2\alpha'p_r\cdot p_s}\;,
\end{equation}
which is a standard result. As with the Virasoro-Shapiro amplitude, this may be cast in the form of $\Gamma$-functions using conformal symmetry or $c$-ghost insertions.

\section{Discussion}

In this Letter, we have extended the NN-FT correspondence to string theory for the first time. Specifically,
we have realized the bosonic string as a NN-FT and utilized this description to provide new computations of famous string amplitudes. The worldsheet embedding $X^\mu$ and the $bc$-ghost system are realized as infinite width limits of neural network architectures with prescribed parameter distributions.
The Veneziano and Virasoro-Shapiro amplitudes are computed as correlators in network parameter space, recovering the standard results from string theory. Spacetime momentum conservation of the amplitudes emerges naturally from translational invariance of the network arising in the infinite variance limit of a Gaussian-distributed zero mode, or by taking the ratio of momentum cutoffs $\Lambda/\epsilon\to\infty$. The string tension scales with the variance of output weights.

What about other neural network architectures? For simplicity we have chosen a fully connected architecture with a single hidden layer, but many other architectures are possible. For instance, for the free boson one simply needs an architecture with 2d input that in the large-$N$ limit is a Gaussian Process with power spectrum $G^{(2)}(p) \sim 1/p^2$. Let
\begin{equation}
    X^\mu(z) = \frac{1}{\sqrt{N}} \sum_i a^\mu_i \varphi_i(z)
\end{equation}
be any such architecture where $\varphi_i$ are neurons of any architecture with $1/p^2$ power spectrum. Take $a^\mu_i$ are i.i.d. across the $i$ index and $\bE[a]=0, \bE[a^2]=:\sigma_a^2$.  
Then   $\langle X^\mu X^\nu\rangle(p) \propto \sigma_a^2 \delta^{\mu\nu}/p^2$ and therefore $\sigma_a$ tunes $\alpha'$. Doing this concretely would  be a different realization of the same GP as \eqref{eq:X^mu_NNFT} as $N\to\infty$, but $1/N$ corrections may induce architecture dependent non-Gaussianities. It would be interesting to study this for different architectures, including corrections to the Veneziano amplitude.

The free boson architecture \cite{Robinson:2025ybg} is closely related to ours, but puts the input-weight dependence that yields the $1/p^2$ in the density, rather than our $1/|W|$ in the architecture \eqref{eq:X^mu_NNFT}. This is a simultaneous rescaling of the architecture and probability density that leaves the GP invariant, a mechanism to generate families of theories with the same GP. The two architectures  have slightly different neurons $\varphi_i$ that are related to a seminal ML result \cite{rahimi2007random} known as  random Fourier features, which have translation-invariant two-point functions. 

Several extensions of this work merit exploration. Already in the bosonic string, it should be possible to realize T-duality, D-branes, and the critical dimension; the latter could follow from central charge techniques of \cite{Robinson:2025ybg}. The RNS superstring can likely be realized by incorporating free fermions and the $\beta\gamma$-ghost system. Finite-width corrections induce interactions in the NN-FT that limits to the bosonic string; their relevance for string theory is unclear. Finally, acting with subsequent network layers $\Phi$ on our first two layers $X^\mu$ defines a deep network $\Phi(X^\mu)$ that might be interesting for string field theory, providing a clear motivation for studying deeper architectures.

\medskip
\noindent \textbf{Acknowledgments.} We thank Christian Ferko, Dan Hackett, and especially Brandon Robinson for discussions and comments. J.H. and S.F. are supported by NSF grant PHY-2209903. J.H. is also supported by the National Science Foundation
under Cooperative Agreement PHY-2019786 (the NSF AI Institute for Artificial Intelligence
and Fundamental Interactions). This work was
supported in part by grant NSF PHY-2309135 to the Kavli Institute for Theoretical Physics (KITP).

\appendix
\onecolumngrid

\section{Virasoro-Shapiro Amplitude}
The Virasoro-Shapiro Amplitude is the tree-level four-point scattering amplitude for closed string tachyons. We define it by:
\begin{equation}
    \mathcal{A}^{(4)}(p_1,\dots,p_4)=\frac{1}{g_s^2}\frac{1}{\text{Vol}(SL(2,\mathbb{C}))}\Big\langle \prod_{i=1}^4 V(p_i)\Big \rangle,\label{eq:VS_defn}
\end{equation}
which depends on the vertex operator
\begin{equation}
    V(p)=\mathcal{Z}(p)\,g_s\int d^2z\,e^{ip\cdot \tilde{X}(z)}\;,
\end{equation}
where $\mathcal{Z}(p)$ is a renormalization factor and $g_s$ is a hyperparameter corresponding to the string coupling.

Using the NNFT correspondence, \eqref{eq:VS_defn} can be rewritten as an integral over neural network parameters:
\begin{align}
    \mathcal{A}^{(4)}(p_1,\dots,p_4)&=\frac{g_s^2}{\text{Vol}(SL(2,\mathbb{C}))}\int [d\mu_X]\;\Big(\prod_{\alpha=1}^4 d^2z_\alpha\Big)\mathcal{Z}(p_1)\dots \mathcal{Z}(p_4)\;\exp\Big(i\sum_{j=1}^4 \sum_{\mu=1}^D\,p_j^\mu X^\mu(z_j)\Big)\notag\\
    &\qquad\qquad\qquad\qquad\qquad\qquad\qquad\qquad\qquad\times\int d^DX_0\,\frac{1}{(2\pi\sigma_0^2)^{D/2}}\;e^{-X_0^2/2\sigma_0^2+i\sum_{j=1}^4 p_j\cdot X_0}\notag\\
    &=\frac{g_s^2\exp[-\frac{\sigma_0^2}{2}(\sum_i p_i)^2]}{\text{Vol}(SL(2,\mathbb{C}))}\int [d\mu_X]\;\Big(\prod_{\alpha=1}^4 d^2z_\alpha\Big)\mathcal{Z}(p_1)\dots \mathcal{Z}(p_4)\;\exp\Big(i\sum_{j=1}^4 \sum_{\mu=1}^D\,p_j^\mu X^\mu(z_j)\Big)\label{eq:VS_amp_params}\;,
\end{align}
where the integration measure $d\mu_X$ is given by:
\begin{equation}
\label{eqn:no_zero_mode_measure}
    d\mu_X=\prod_{i=1}^N \bigg[\frac{d^2 W_i}{\text{Vol}(B^2_\Lambda\backslash B^2_\epsilon)}\,\frac{dc_i}{2\pi}\,\prod_{\mu=1}^D \frac{da_i^\mu}{\sqrt{2\pi\sigma_a^2}}\,e^{-(a_i^\mu)^2/2\sigma_a^2}\bigg]\;.
\end{equation}
We now compute expectation with respect to $a$. 
Defining $A:=\frac{1}{\sigma_a}\sqrt{\frac{\alpha'(\Lambda^2-\epsilon^2)}{N}}$ and $f_i(z):=\frac{\cos[(W_i z+\bar{W}_i\bar{z})/2+c_i]}{|W_i|}$ such that the integrand of \eqref{eq:VS_amp_params} (ignoring the renormalization factors for now) is:
\begin{equation}
    \exp\Big(iA\sum_{i=1}^N \sum_{\mu=1}^D \sum_{j=1}^4 a_i^\mu\,p_j^\mu \,f_i(z_j)\Big)=\exp\Big(iA\sum_{i=1}^N \sum_{\mu=1}^D a_i^\mu S_i^\mu\Big)\;,
\end{equation}
with $S_i^\mu:=\sum_{j=1}^4\, p^\mu_j\,f_i(z_j)$. Then, the $a$ integral becomes:
\begin{align}
    \int \prod_{i,\mu}\Big[\frac{da_i^\mu}{\sqrt{2\pi\sigma_a^2}}e^{-(a_i^\mu)^2/2\sigma_a^2}\,e^{iAa_i^\mu S_i^\mu}\Big]&=\prod_{i,\mu}\int\frac{da}{\sqrt{2\pi\sigma_a^2}}e^{-a^2/2\sigma_a^2}\,e^{iAa S_i^\mu}\notag\\
    &=\exp\Big[-\frac{1}{2}A^2 \sigma_a^2 \sum_{i=1}^N \sum_{\mu=1}^D (S_i^\mu)^2\Big]\notag\\
    &=\exp\Big[-\frac{1}{2}A^2\sigma_a^2\sum_{i=1}^N \sum_{r,s=1}^4 (p_r\cdot p_s)\,f_i(z_r) f_i(z_s)\Big]\notag\\[10 pt]
    &=\exp\Big[-\frac{\alpha'(\Lambda^2-\epsilon^2)}{2N}\sum_{i=1}^N \sum_{r,s=1}^4 (p_r\cdot p_s)\,f_i(z_r) f_i(z_s)\Big]\label{eq:a_int_result}\;.
\end{align}

It is now convenient to define an intermediate variable $Q$ by:
\begin{equation}
    Q:=\sum_{r,s=1}^4 (p_r\cdot p_s)f_i(z_r) f_i(z_s)=\sum_{r,s=1}^4 (p_r\cdot p_s)\frac{\cos\Big(\frac{1}{2}(W z_r+\bar{W}\bar{z}_r)+c\Big)\cos\Big(\frac{1}{2}(W z_s+\bar{W}\bar{z}_s)+c\Big)}{|W|^2}\notag\;,
\end{equation}
such that the Virasoro-Shapiro amplitude is
\begin{align}
    \mathcal{A}^{(4)}(p_1,\dots,p_4)&=\frac{g_s^2\,e^{-\sigma_0^2 (\sum_i p_i)^2/2}}{\text{Vol}(SL(2,\mathbb{C}))}\int \Big(\prod_{\alpha=1}^4 d^2z_\alpha\Big)\;\mathcal{Z}(p_1)\dots\mathcal{Z}(p_4)\,\mathbb{E}_{W,c}\Big[\exp\Big(-\frac{\alpha'(\Lambda^2-\epsilon^2)}{2N}\,Q\Big)\Big]^N\notag\\[10 pt]
    &:=\frac{g_s^2\,e^{-\sigma_0^2 (\sum_i p_i)^2/2}}{\text{Vol}(SL(2,\mathbb{C}))}\int \Big(\prod_{\alpha=1}^4 d^2x_\alpha\Big)\;\mathcal{Z}(p_1)\dots\mathcal{Z}(p_4)\,[\Phi_N]^N\label{eq:VS_intermediate}\;,
\end{align}
where we have used that the $W$ and $c$ parameters are drawn i.i.d across the $i$ index. By taking the Gaussian limit $N\to\infty$, the integrand of \eqref{eq:VS_intermediate} is, to leading order in $1/N$,
\begin{equation}
    [\Phi_N]^N=\exp[N\log\Phi_N]=\exp\Big(-\frac{\alpha'}{2}(\Lambda^2-\epsilon^2)\,\mathbb{E}_{W,c}[Q]\Big)\notag\;.
\end{equation}
The expectation $\mathbb{E}_c[Q]$ is simple because $\mathbb{E}_c[\cos(y+c)]=0$ for all $y$ allows a simplification at an intermediate step. Special care must be taken to separate the sum $\sum_{r,s}=\sum_{r\neq s}+\sum_r$ before taking expectation with respect to $W$:
\begin{align}
    \mathbb{E}_{W,c}[Q]&=\mathbb{E}_W\Big[\frac{1}{2}\sum_{r,s=1}^4 (p_r\cdot p_s)\,\frac{\cos\big[(W\Delta z_{rs}+\bar{W}\Delta \bar{z}_{rs})/2\big]}{|W|^2}\Big]\notag\\[10 pt]
    &=\mathbb{E}_W\Big[\frac{1}{2}\sum_{r\neq s}\,(p_r\cdot p_s)\,\frac{\cos\big[(W\Delta z_{rs}+\bar{W}\Delta \bar{z}_{rs})/2\big]}{|W|^2}+\frac{1}{2}\sum_r\,p_r^2\,\frac{1}{|W|^2}\Big]\;\label{eq:expec_Q}.
\end{align}
The first term in \eqref{eq:expec_Q} gives the standard boson propagator; the second is a trivial logarithmic integral:
\begin{align}
    \mathbb{E}_{W,c}[Q]&=\frac{1}{\Lambda^2-\epsilon^2}\sum_{r\neq s}\,(p_r\cdot p_s)\,\Big[-\log|z_r-z_s|+\log\Big(\frac{2\Lambda}{\epsilon e^\gamma}\Big)\Big]+\frac{1}{\Lambda^2-\epsilon^2}\sum_r p_r^2\,\ln\Big(\frac{\Lambda}{\epsilon}\Big)\notag\\
    &=-\frac{1}{\Lambda^2-\epsilon^2}\sum_{r\neq s}(p_r\cdot p_s)\log|z_r-z_s|+C(\Lambda,\epsilon,p)\;,
\end{align}
where we have defined the cutoff-dependent constant:
\begin{equation}
    C(\Lambda,\epsilon,p)=\frac{1}{\Lambda^2-\epsilon^2}\Big[\Big(\sum_r p_r\Big)^2\log\frac{2\Lambda}{\epsilon e^\gamma}-\sum_r p_r^2 \log\frac{2}{e^\gamma}\Big]\;.
\end{equation}
Then, the integrand of \eqref{eq:VS_intermediate} up to renormalization factors is
\begin{equation}
    \exp\Big(-\frac{\alpha'(\Lambda^2-\epsilon^2)}{2}\mathbb{E}_{W,c}[Q]\Big)=\exp\Big(\frac{\alpha'}{2}\sum_{r\neq s}(p_r\cdot p_s)\log|z_r-z_s|\Big)\exp\Big(-\frac{\alpha'}{2}\log\frac{2\Lambda}{\epsilon e^\gamma}\big(\sum_r p_r\big)^2\Big)\exp\Big(\frac{\alpha'}{2}\sum_r p_r^2 \log\frac{2}{e^\gamma}\Big)\;.
\end{equation}
We will look at each of these exponential factors individually. The first is simplified as
\begin{align}
    \exp\Big(\frac{\alpha'}{2}\sum_{r\neq s}(p_r\cdot p_s)\log|z_r-z_s|\Big)&=\exp\Big[\log\Big(\prod_{r\neq s}|z_r-z_s|^{\alpha'p_r\cdot p_s/2}\Big)\Big]\notag\\
    &=\prod_{r<s}|z_r-z_s|^{\alpha'p_r\cdot p_s}\;.
\end{align}
The second factor combines with the Gaussian involving $\sigma_0^2$ outside the integral to define a new Gaussian $\exp[-\sigma_{\text{eff}}^2(\sum_r p_r)^2/2]$ with (inverse) variance given by
\begin{equation}
    \sigma_{\text{eff}}^2=\sigma_0^2+\alpha'\log\frac{2\Lambda}{\epsilon e^\gamma}\label{eq:new_gaussian}\;.
\end{equation}
The third exponential factor simplifies to:
\begin{equation}
    \exp\Big(\frac{\alpha'}{2}\log\frac{2}{e^\gamma}\,\sum_{r=1}^4 p_r^2\Big)=\prod_{r=1}^4 \exp\Big(\frac{\alpha'}{2} p_r^2\log\frac{2}{e^\gamma}\Big)\label{eq:p^2_terms}\;.
\end{equation}
These terms come precisely from the $r=s$ case and would not arise if we had utilized normally ordered vertex operators. This may be effectively enforced by choosing renormalization factors
$\mathcal{Z}(p)$ such that they cancel with \eqref{eq:p^2_terms}. 
\begin{equation}
    \mathcal{Z}(p)=\Big(\frac{e^\gamma}{2}\Big)^{\alpha' p^2/2}\,\Big(\frac{\sigma^2_{\text{eff}}}{2\pi}\Big)^{D/8}\;.
\end{equation}
This choice also provides the normalization for the Gaussian that becomes the $\delta$-function below.
Putting it all together, the final result is:
\begin{equation}
    \mathcal{A}^{(4)}=\frac{g_s^2\left(\frac{\sigma^2_\text{eff}}{2\pi}\right)^{\frac{D}{2}}e^{-\frac{\sigma_{\text{eff}}^2(\sum_r p_r)^2}{2}}}{\text{Vol}(SL(2,\mathbb{C}))}
    \int\Big(\prod_{\alpha=1}^4 d^2z_\alpha\Big)\,\prod_{r<s}|z_r-z_s|^{\alpha'p_r\cdot p_s}\;.
\end{equation}
Taking the $\sigma_{\text{eff}}\to\infty$ limit,
\begin{equation}
    \mathcal{A}^{(4)}=\frac{g_s^2\,\delta^{D}(\sum_r p_r)}{\text{Vol}(SL(2,\mathbb{C}))}\int\Big(\prod_{\alpha=1}^4 d^2z_\alpha\Big)\,\prod_{r<s}|z_r-z_s|^{\alpha'p_r\cdot p_s}\;,
\end{equation}
which is the standard result for the Virasoro-Shapiro amplitude; see, e.g., (6.9) of \cite{Tong:2009np}. The volume factor may be removed by using conformal symmetry to fix three of the $z$'s, leaving the standard result in terms of $\Gamma$-functions.

\section{Veneziano Amplitude}
The Veneziano Amplitude is the tree-level four-point scattering amplitude for open string tachyons. We define it by:
\begin{equation}
    \mathcal{A}^{(4)}(p_1,\dots,p_4)=\frac{1}{g_s}\frac{1}{\text{Vol}(SL(2,\mathbb{R}))}\Big\langle\prod_{i=1}^4 V_o(p_i)\Big\rangle\;,
\end{equation}
where the open string vertex operator
\begin{equation}
    V_o(p)=\mathcal{Z}_o(p)\,\sqrt{g_s}\int dx\,e^{ip\cdot\tilde{X}_N(x)}\notag
\end{equation}
depends on renormalization factors $\mathcal{Z}_o(p)$ and the open string $\tilde X_N(x)$ with Neumann boundary conditions evaluated at $x\in\mathbb{R}$. This architecture is given by:
\begin{equation}
    \tilde X^\mu_N(x)=C\sqrt{\frac{2}{N}}\sum_{i=1}^N \frac{a_i^\mu}{|W_i|}\cos\Big(\frac{1}{2}(W_i+\bar{W}_i)x+c_i\Big)+X^\mu_0\label{eq:open_boundary_arch}\;,
\end{equation}
where $X_0^\mu$ is a Gaussian distributed zero mode. With these modifications, the calculation of the Veneziano Amplitude mirrors that of the Virasoro-Shapiro Amplitude. The integral over $X_0$ is the same. The extra factor of $\sqrt{2}$ in the architecture \eqref{eq:open_boundary_arch} means that the integral over $a$ parameters gives the same answer as in \eqref{eq:a_int_result} but without the factor of 2 in the denominator of the exponent. Then, the calculation of $\mathbb{E}_{W,c}[Q]$ proceeds identically, so the end result is that relative to the Virasoro-Shapiro case, there are three places in the final answer where a factor of 2 shows up. One is in the exponent of $|x_r-x_s|$. Another is in the definition of a new ${\sigma'}^2_{\text{eff}}=\sigma_0^2+2\alpha'\log(2\Lambda/\epsilon e^\gamma)$. The third is in the renormalization factors:
\begin{equation}
    \mathcal{Z}_o(p)=\Big(\frac{e^\gamma}{2}\Big)^{\alpha' p^2}\Big(\frac{{\sigma'}_{\text{eff}}^2}{2\pi}\Big)^{D/8}\;,
\end{equation}
where the exponents are $\alpha' p^2$ instead of $\alpha' p^2/2$. The final expression for the Veneziano Amplitude is therefore:
\begin{equation}
    \mathcal{A}^{(4)}=\frac{g_s\left(\frac{{\sigma'}^2_\text{eff}}{2\pi}\right)^{\frac{D}{2}}e^{-\frac{{\sigma'}_{\text{eff}}^2(\sum_r p_r)^2}{2}}}{\text{Vol}(SL(2,\mathbb{R}))}
    \int\Big(\prod_{\alpha=1}^4 dx_\alpha\Big)\,\prod_{r<s}|x_r-x_s|^{2\alpha'p_r\cdot p_s}\;,
\end{equation}
and taking $\sigma'_{\text{eff}}\to\infty$:
\begin{equation}
    \mathcal{A}^{(4)}=\frac{g_s\,\delta^{D}(\sum_r p_r)}{\text{Vol}(SL(2,\mathbb{R}))}\int\Big(\prod_{\alpha=1}^4 dx_\alpha\Big)\,\prod_{r<s}|x_r-x_s|^{2\alpha'p_r\cdot p_s}\;,
\end{equation}
recovering the standard result.

\bibliography{refs}

\end{document}